# Simulating Plasmon Resonances of Gold Nanoparticles with Bipyramidal Shapes by Boundary Element Methods

Jacopo Marcheselli, Denis Chateau, Frederic Lerouge, Patrice Baldeck, Chantal Andraud, Stephane Parola, Stefano Baroni,* Stefano Corni,* Marco Garavelli,* and Ivan Rivalta*

ACCESS | Metrics & More | Article Recommendations | Supporting Information

**ABSTRACT:** Computational modeling and accurate simulations of localized surface plasmon resonance (LSPR) absorption properties are reported for gold nanobipyramids (GNBs), a class of metal nanoparticle that features highly tunable, geometry-dependent optical properties. GNB bicone models with spherical tips performed best in reproducing experimental LSPR spectra while the comparison with other geometrical models provided a fundamental understanding of base shapes and tip effects on the optical properties of GNBs. Our results demonstrated the importance of averaging all geometrical parameters determined from transmission electron microscopy images to build representative models of GNBs. By assessing the performances of LSPR absorption spectra simulations based on a quasi-static approximation, we provided an applicability range of this approach as a function of the nanoparticle size, paving the way to the theoretical study of the coupling between molecular electron densities and metal nanoparticles in GNB-based nanohybrid systems, with potential applications in the design of nanomaterials for bioimaging, optics and photocatalysis.

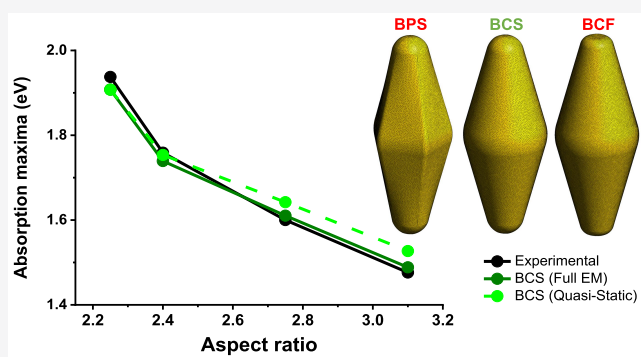

## INTRODUCTION

Metal nanoparticles, noble metals in particular, feature a wide spectral tuning of the localized surface plasmon resonance (LSPR),[1,2] as a function of their size, shape, and dielectric environment.[3−6] Gold nanoparticles (NPs) with bipyramidal shape (gold nanobipyramids, GNBs, see Figure 1) are particularly well-known for their high tunability and optical sensitivity,[7] featuring a narrower plasmon band with respect to other nonspherical nanoparticles, like rods or dog-bone[8−10] and allowing fine-tuning of the plasmon resonance energy.[11−17] Moreover, the tips of gold nanobipyramids enhance the effects of the local electric field generated by the metal nanoparticle, which make these nano-objects particularly sensitive to local changes in the dielectric environment, suitable even for detection of single molecules[18,19] and various technological applications. In particular, given the possibility of producing GNBs with variable size and aspect ratio (AR) at high yields and in almost monodispersed samples,[20,21] these nano-objects represent ideal inorganic partners for the assembly of hybrid organic−inorganic nanoparticles with peculiar optical properties, with potential impact in technological devices for bioimaging, photocatalysis, optics, solar cells, and biotechnology.[22−24]

The interaction between the optically induced collective excitations of free charges confined in the metal nanoparticle surface and the excitons of organic chromophores is one of the main features of metal−organic nanohybrids.[25] Moreover, upon resonant excitation, LSPR can exceed the diffraction limit and collect light into a subwavelength region, greatly enhancing the electric field acting over optically active molecules adjacent (adsorbed or physisorbed) to the metal surface.[26−28] In general, the LSPR operates quenching or enhancing the photophysical properties of the nearby molecular systems, resulting for instance in modified absorbance and/or fluorescence, or in preserving the photochemical stability of the chromophores by opening nonreactive excited state deactivation processes.[14,29] To allow and control such intriguing synergistic optical properties and to design new nanomaterials for versatile technological applications, computational modeling could be extremely beneficial, as it could provide rationalization of experimental evidence and eventually in silico prediction of relevant properties. However, computing in details the physicochemical properties of metal−organic nanohybrids proves to be a difficult computational challenge. While the nature of the organic component requires a full

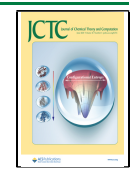

3807

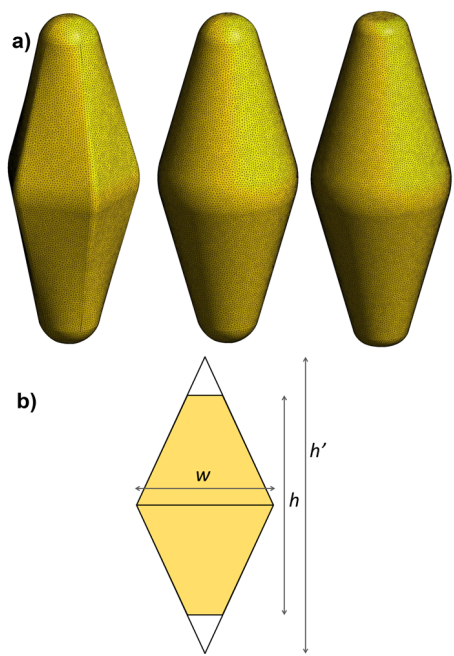

**Figure 1.** (a) The GNBs' models used in this work, including, from left to right the bipyramidal with pentagonal base and spherical tip (BPS), the biconical with spherical tip (BCS) and the biconical with flat tip (BCF) shapes. (b) A schematic representation of the basic geometric parameters of GNBs, with base length ($w$), height ($h$), and 2D projected area (in yellow) being extracted from TEM images, from which the nontruncated, ideal height ($h'$) can be derived.

quantum-mechanical (QM) treatment in order to be predictive and properly interpret molecular electronic properties, the size of the inorganic component renders the QM treatment computationally too expensive, hampering the employment of an homogeneous technique for supramolecular nanohybrid systems. For this reason, hybrid models have been developed combining a QM level treatment and a classical electromagnetism approach for molecules and the inorganic nanostructures, respectively.[30] In particular, state-of-the-art methodologies[31−35] treat the inorganic metallic component within a polarizable continuum model (PCM) approach,[36−38] i.e., as a continuous body characterized by its own frequency dependent dielectric function, while the molecular excitations can be determined by employing the time-dependent density functional theory (TDDFT).[39] Still, such hybrid computational modeling has not been reported for nanohybrids containing gold bipyramids, despite their relevant optical properties, their potential application as inorganic components in nanohybrid materials and their use as seeds for regrowing nanoparticles with various shapes.[11,20] The first theoretical challenge is to achieve an accurate and comprehensive modeling of pure GNBs, with previously reported simulations possessing limited accuracy both in describing structural details of various bipyramidal morphologies (e.g., smooth edges and tips) and in reproducing experimental LSPR absorption maxima as a function of the GNB's AR.[12,14]

The theoretical treatment of LSPR in noble metal nanoparticles has long been established,[1,40] with methods that generally stem from classical electrodynamics Mie scattering theory, which gives analytical results only for simple spheres, such as spheres, imposing the use of numerical approaches for other shapes, like bipyramids. These methods generally include modeling of the environment around the metallic object as a continuum dielectric, usually proving to be effective in the description of LSPR spectra and near-field enhancements, in the interpretation of experimental observations[41] and in the design new materials for technological applications.[42] For GNBs, the discrete dipole approximation (DDA)[43] has been used to solve numerically the electromagnetic (EM) scattering equations[12,14] for a nano-object, where the three-dimensional solid particle is replaced by a finite array of polarizable point dipoles, with the spacing between the dipoles small compared to the wavelength of the incoming light. Each of these dipoles has an oscillating polarization in response to both an incident plane wave and the electric fields due to all of the other dipoles in the array. This numerical method allows the employment of arbitrary shapes, instead of being restricted to a small number of regular shapes like spheres, spheroids and infinite cylinders, along with the opportunity to introduce anisotropies inside the particle. However, this approach has some limitations in terms of affordability with respect to the number of dipoles employed, which does not allow for a very detailed representation of complex shapes. An alternative approach is to solve numerically the EM scattering equations exploiting the boundary element method (BEM). BEM, in general, is a numerical computational method of solving linear partial differential equations which were reformulated as integral equations. Rather than solving a numerical problem of a set of partial differential equations defined throughout the whole space, BEM maps the problem into an integral equation defined over the boundary of the system. In practice, a three-dimensional problem (like that solved in DDA method) is mapped onto a two-dimensional surface mesh, thus demanding a computational cost that scales with the size of the surface instead of the volume and then allowing for a greater control of the details of the chosen object. The solid particle is represented by a set of apparent surface charges and currents that can be used to solve EM scattering equations and compute the absorption and scattering of the dielectric object.[44−47] In particular, the metal response to the electric fields can be determined using the widespread assumption that the metal behaves as a dielectric for time dependent fields. Within this approach, the solvent can still be described as a continuum dielectric which occupies all the space free from the metal specimen.

In this work, we report the unprecedented modeling of GNBs by means of the BEM approach, comparing the resulting optical properties with DDA results and with a large set of experimental data, and we introduce a BEM-based modeling for future studies of organic−inorganic nanohybrids based on GNBs. We first focused on providing a comprehensive theoretical investigation of GNBs to link the LSPR absorption maxima to their morphological characteristics, i.e., various kinds of tips or possible base geometries, even when a distribution of GNBs (as synthetic nanoparticles usually appear in solution) is taken into account, thus determining the effect of the statistical distribution of geometrical parameters on the LSPR absorption of GNBs with respect to single GNBs. If the size of the NP is comparable to the size of the incident wave, then the full-EM equation is needed. If the size of the NP is much smaller than the incident wave, the full-EM (i.e., Maxwell) equations reduce to just a time dependent Poisson problem. This approximation is called quasi-static (or dipole) approximation (QSA), and it allows straightforwardly to couple organic chromophores described by first principle approaches with the metal NP.[30] However, within this approximation the particle size does not affect the computed LSPR absorption energy but only its intensity, making the QSA simulations of LSPR absorption energies independent from the particle sizes. Here, we employed the BEM approach within the QSA to determine the maximum size of the GNBs that can be modeled with reasonable accuracy, paving the way to the study of GNB-based nanohybrids optical properties.

## ■ COMPUTATIONAL DETAILS

Numerical computations of LSPR absorption maxima have been performed using two different codes: Scuff-em, a free, open-source software implementation of the BEM method developed by Reid et al.[47,48] This code solves the full-EM scattering equations, taking into account explicitly the size of the GNBs; and TDPLAS,[49] that is routinely used to compute interaction between molecular electron densities and metal NPs, and presently relies on the QSA approximation.

The BEM approach implemented in Scuff-em makes use of effective (or apparent) electric ($\mathbf{K}$) and magnetic ($\mathbf{N}$) current densities located on the boundary of the nanoparticle and tangential to it to express the EM fields (electric and magnetic field) everywhere in space. By imposing the proper EM boundary conditions at the NP surfaces, it turns out that such current densities should satisfy the linear equation:

$$
\begin{aligned}
-\mathbf{E}_{\parallel}^{\text{inc}}(x) &= \left[\oint_{\partial\Omega_r} \{\mathbf{\Gamma}^{EE,r}\cdot\mathbf{K} + \mathbf{\Gamma}^{EM,r}\cdot\mathbf{N}\}\,dx' \right.\\
&\quad \left. - \oint_{\partial\Omega_s} \{\mathbf{\Gamma}^{EE,s}\cdot\mathbf{K} + \mathbf{\Gamma}^{EM,s}\cdot\mathbf{N}\}\,dx' \right]_{\parallel}\\
-\mathbf{H}_{\parallel}^{\text{inc}}(x) &= \left[\oint_{\partial\Omega_r} \{\mathbf{\Gamma}^{ME,r}\cdot\mathbf{K} + \mathbf{\Gamma}^{MM,r}\cdot\mathbf{N}\}\,dx' \right.\\
&\quad \left. - \oint_{\partial\Omega_s} \{\mathbf{\Gamma}^{ME,s}\cdot\mathbf{K} + \mathbf{\Gamma}^{MM,s}\cdot\mathbf{N}\}\,dx' \right]_{\parallel}
\end{aligned}
\quad (1)
$$

where the integrals are on the NP boundary from the inside ($\partial\Omega_r$) and from the outside ($\partial\Omega_s$) of the NP and $\Gamma^{XX,t}$ is the proper electromagnetic dyadic Green's function ($X$=Electric or Magnetic) for the interior ($t = r$) or the exterior ($t = s$) of the NP evaluated at the incident EM field frequency. $\Gamma^{XX,t}$ contains the dependence on the NP dielectric function $\epsilon(\omega)$ (as well as on the magnetic susceptibility $\mu(\omega)$, that however we take equal to the vacuum susceptibility $\mu_0$ for all frequencies since gold is not a magnetic material); we omit dependence on frequency in the formulas for clarity. The expressions of $\Gamma^{XX,t}$ can be found in ref 50. $\mathbf{E}_{\parallel}^{\text{inc}}$ and $\mathbf{H}_{\parallel}^{\text{inc}}$ in eq 1 are the components tangential ($\parallel$) to the NP surface of the incident electric and magnetic fields. To numerically solve eq 1, the surface integrals are discretized in finite boundary elements, and the surface current densities $\mathbf{K}$, $\mathbf{N}$ are expanded on a finite basis set:

$$
\mathbf{K}(x) = \sum_{\alpha} k_{\alpha}\mathbf{f}_{\alpha}(x), \quad \mathbf{N}(x) = -Z_0\sum_{\alpha} k_{\alpha}\mathbf{f}_{\alpha}(x) \quad (2)
$$

Converting eq 1 in a standard linear equation problem:

$$
\begin{pmatrix} \mathbf{M}^{EE} & \mathbf{M}^{EM} \\ \mathbf{M}^{ME} & \mathbf{M}^{MM} \end{pmatrix} \cdot \begin{pmatrix} \mathbf{K} \\ \mathbf{N} \end{pmatrix} = \begin{pmatrix} \mathbf{V}^E \\ \mathbf{V}^N \end{pmatrix}
\quad (3)
$$

with

$$
\mathbf{V}_{\alpha}^E = -\langle\mathbf{f}_{\alpha}|\mathbf{E}^{\text{inc}}\rangle/Z_0, \quad \mathbf{V}_{\alpha}^M = -\langle\mathbf{f}_{\alpha}|\mathbf{H}^{\text{inc}}\rangle \quad (4)
$$

and

$$
\begin{aligned}
\mathbf{M}_{\alpha\beta}^{EE}(\xi) &= \langle\mathbf{f}_{\alpha}|\mathbf{\Gamma}^{EE,s}(\xi) + \mathbf{\Gamma}^{EE,r}(\xi)|\mathbf{f}_{\beta}\rangle/Z_0 \\
\mathbf{M}_{\alpha\beta}^{EM}(\xi) &= -\langle\mathbf{f}_{\alpha}|\mathbf{\Gamma}^{EM,s}(\xi) + \mathbf{\Gamma}^{EM,r}(\xi)|\mathbf{f}_{\beta}\rangle \\
\mathbf{M}_{\alpha\beta}^{ME}(\xi) &= \langle\mathbf{f}_{\alpha}|\mathbf{\Gamma}^{ME,s}(\xi) + \mathbf{\Gamma}^{ME,r}(\xi)|\mathbf{f}_{\beta}\rangle \\
\mathbf{M}_{\alpha\beta}^{MM}(\xi) &= -Z_0\langle\mathbf{f}_{\alpha}|\mathbf{\Gamma}^{MM,s}(\xi) + \mathbf{\Gamma}^{MM,r}(\xi)|\mathbf{f}_{\beta}\rangle
\end{aligned}
\quad (5)
$$

$Z_0$ is the impedance of free space and $|\mathbf{f}_{\alpha}\rangle$ is a vector of basis set elements that are by construction tangent to the surface. Scuff-em makes use in particular of the RWG basis set.[51] Once the vectors $\mathbf{K}$ and $\mathbf{N}$ have been obtained, they can be used to calculate the EM field power absorbed by the NP ($P^{\text{abs}}$) via the calculation of the Poynting vector flux through the NP surface:

$$
P^{\text{abs}} = \frac{1}{2}\mathcal{R}\oint_{\partial\Omega} \mathbf{K}^*(x)\cdot[\hat{n} \times N(x)]\,dx \quad (6)
$$

TDPLAS exploits the QSA where it is possible to simplify the full EM BEM eq 1. In particular, the theory can be reformulated in terms of apparent charge densities (rather than currents). An account on how to move from full EM BEM equations to QSA ones is given, e.g., in ref 52. In TDPlas implementation, the surface charge density is assumed to be constant on each boundary element $i$, giving rise to a total charge $q_i$. Such charges are also obtained by solving a linear equation leading to[49]

$$
\mathbf{q}(\omega) = \mathbf{Q}(\omega)\mathbf{g}(\omega) \quad (7)
$$

where $g_i(\omega)$ is given by $-\mathbf{E}^{\text{inc}}\cdot\mathbf{s}_i$ ($\mathbf{s}_i$ is the center of the boundary element $i$) and:

$$
\mathbf{Q} = -\mathbf{S}^{-1}\left(2\pi\frac{\epsilon(\omega) + 1}{\epsilon(\omega) - 1}\mathbf{I} + \mathbf{D}\mathbf{A}\right)^{-1}(2\pi\mathbf{I} + \mathbf{D}\mathbf{A}) \quad (8)
$$

is the NP response matrix containing the dielectric function $\epsilon(\omega)$, the diagonal matrix $\mathbf{A}$ collects the areas of the boundary elements and the $\mathbf{S}$ and $\mathbf{D}$ matrices are related to electric field and electric potential integrals:

$$
D_{ij} = \frac{(\mathbf{s}_i - \mathbf{s}_j)\cdot\mathbf{n}_j}{|\mathbf{s}_i - \mathbf{s}_j|^3}, \quad S_{ij} = \frac{1}{|\mathbf{s}_i - \mathbf{s}_j|} \quad (9)
$$

for out of diagonal elements (for diagonal elements see ref 31). $\mathbf{n}_j$ is the unit vector normal to the boundary element $j$ and pointing outward. Since $\epsilon(\omega)$ is complex, also $\mathbf{q}$ are complex. Once the charges $\mathbf{q}$ are obtained for a unit electric field set along a given direction $\mathbf{e}$, a polarizability elements $\alpha_{fe}$ can be obtained as $\alpha_{fe} = \sum f_i q_i$ and the overall absorbed power at the frequency $\omega$ can be obtained in atomic units as $4\pi\omega/c\cdot\mathcal{J}[Tr(\alpha)]$ where $\mathcal{J}$ is the imaginary part and $Tr$ is the trace of.

GNBs' modeling is based on experimental geometrical parameters extracted from transmission electron microscopy (TEM) data of synthetic GNBs, as reported by Chateau et al.,[12] Sánchez-Iglesias et al.,[20] and Chateau et al.[21] Among the GNBs synthesized by Chateau et al.[12] we have considered the NPs for which the DDA simulated and the experimental LSPR maxima were reported and whose volumes were explicitly indicated, as determined from TEM images. For the GNBs synthesized by Sánchez-Iglesias et al.,[20] we selected the samples where the NPs appeared less aggregated, allowing a more accurate image analysis of the TEM pictures (i.e., the GNBs reported in Figure S10a−c in the Supporting Information of ref 20).

The analysis of TEM images show that the selected GNBs could have either a pentagonal (bipyramids) or a spherical (bicones) base shape. Indeed, the penta-twinned seeds used for the synthesis of GNBs have pentagonal base but GNBs' edges are generally strongly smoothed, as observed in 3D tomographic electron microscopy at the single-particle level and in high-resolution TEM images.[11,16,53] Therefore, three GNB model shapes have been here considered: (i) bipyramid with pentagonal base and spherical tip (bipyramid spherical, BPS); (ii) bicone with spherical tip (bicone spherical, BCS); and (iii) bicone with flat tip (bicone flat, BCF). The model shapes and their meshes, built using the code GMSH,[54] are shown in Figure 1a. The smoothed edges are necessary to comply with the BEM method requirement of continuously differentiable surfaces and to minimize as much as possible the nonphysical effects arising from infinitely sharp edges, providing realistic descriptions of tip or edge effects. While for the GNBs synthesized by Chateau et al.[12] the shape (biconical), the $AR$ (including the truncation level from ideal bipyramids), and the size of GNBs (in terms of volume) were explicitly reported, for the remaining GNBs, these parameters were extracted from the TEM images using the image analysis tools provided by ImageJ.[55] The GNBs models have been thus determined, for all shapes, from three experimental geometrical parameters for each single NP: (i) the width of the bipyramidal base ($w$); (ii) the $AR$ ($h/w$), where $w$ and height ($h$) of single particles were considered to be the Feret's minimum and maximum diameter, respectively;[56] and (iii) the "ideal" $AR$ ($h'/w$), namely $AR_{Id}$, as computed by approximating the 2D projection of the GNB to a truncated rhombus (or the union of two symmetric isosceles trapezoids), as depicted in Figure 1b. By simple geometrical considerations, the $AR_{Id}$ can be rewritten as a function of parameters that can be directly extracted from TEM images, reading

$$AR_{Id} = \frac{h^2}{2(h \cdot w - A)}$$

where $A$ is the 2D projected area of the GNBs as obtained from ImageJ analysis of the TEM images. To build the GNB model of a given sample (associated with an experimental LSPR spectrum), thus, we extracted from its TEM image the $w$, $h$ and $A$ values for each NP, allowing computing the average $\langle w \rangle$, $\langle AR \rangle$, and $\langle AR_{Id} \rangle$ over the full set of NPs present in the sample. These three averaged parameters allow the construction of an average GNB structure for each sample. The standard deviations of these structural parameters, i.e., $\sigma(\langle w \rangle)$, $\sigma(\langle AR \rangle)$, and $\sigma(\langle AR_{Id} \rangle)$, were also computed, providing a way to estimate the geometrical uncertainty over the simulated LSPR values, due to the use of a single, average GNB structure as representative of each sample. A full list of the GNBs's parameters is reported in the SI, see Table S2. Finally, for the dielectric properties of gold, we employed the analytic approach reported in ref 57 and 58.

## RESULTS AND DISCUSSION

**Assessment of GNBs' Model Shapes.** Figure 2a illustrates the effects of $AR$ and base size on the LSPR absorption maxima of GNBs particles with the same $AR_{Id}$ and shows the comparison between experimental and theoretical values computed solving the full-EM scattering equations, i.e., without limitation in accuracy due to the specific particle sizes and light wavelengths. Previously reported experimental data and DDA simulations[12] are compared with our BEM results, computed with the Scuff-em code for the three model shapes (i.e., BPS, BCS and BCF) here considered (see Computational Details section). First, we point out that the DDA results (obtained with a biconical shape) match experimental LSPR absorption maxima only for larger $AR$s, while at smaller $AR$s the absolute errors increase and also the trend is not perfectly reproduced (see Figure 2a and Table S1 in the SI). This behavior is most likely due to the limitations of DDA approach in defining with great details the GNB shape (due to the computational cost of the 3D mapping), which become marginal only when the $AR$ and the size of the NP increase significantly. Our BEM computations, instead, do not suffer of such limitations and indeed they provide excellent agreement with experimental data even at small $AR$s if the BCS model, i.e., with round base and spherical tips, is chosen. Notably, if the pentagonal base is considered instead, i.e., the BPS model, the BEM simulations reproduce properly the experimental trends but the LSPR maxima are offset by a consistent and systematic red-shift, always larger than 0.07 eV (see Table S1 in the SI). This systematic red-shift with respect to the biconical BCS shape is due to the presence of the extended lateral edges associated with the pentagonal base, an effect that is clearly out-weighting the (small) blue-shift one should expect when the overall particle volume reduces by passing from a biconical to a bipyramidal shape. In fact, the decrease of metal NPs size is generally accompanied by a blue-shift of the plasmonic resonance absorption.

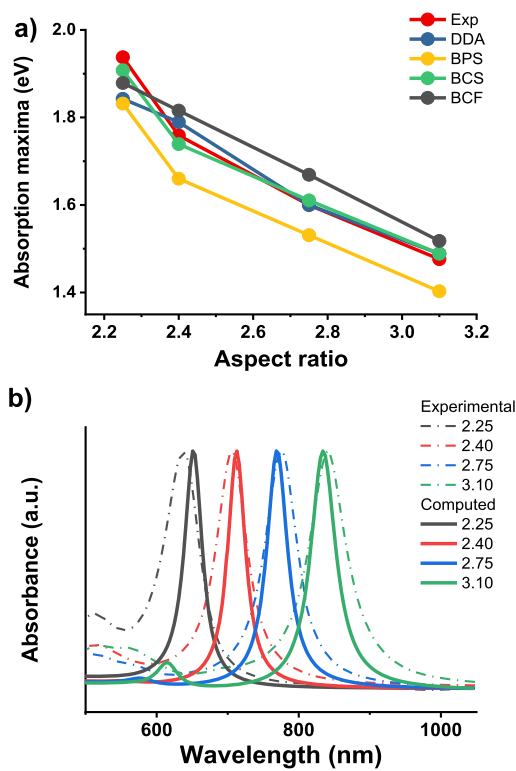

Figure 2. (a) Effects of $AR$ on the LSPR absorption maxima of GNBs particles. Experimental data[12] (in red) are compared with computed values solving the full-EM equations, including previously reported[12] DDA simulations (in blue), and our BEM results adopting the BPS (in yellow), the BCS (in green), and the BCF (in violet) models. (b) Experimental (normalized) absorption electronic spectra of GNBs featuring the various $AR$s reported in (a), including $AR = 2.25$ (in dark gray), $AR = 2.40$ (in red), $AR = 2.75$ (in blue), and $AR = 3.10$ (in green), are compared with spectra computed by our BEM approach using the BCS model.

By comparing the two biconical models, BCS and BCF, it is possible to define the effect of flattening the GNB's tip, which showed to feature two main contributions: (i) the increase of volume particle passing from a spherical to a flat tip for GNBs, which blue-shifts the BCS's LSPR absorption with respect to BCF and (ii) the effect of surface sharpening due to the presence of tips, which induces a red-shift of the plasmonic resonance. As shown in Figure 2a, when the GNB with the smallest $AR$ is considered (i.e., $AR$ = 2.25), which also features the smallest base size (i.e., $w$ = 11 nm), the volume variation due to change of tip shape is significant if related to the overall, relatively small particle volume, indicating the former "volume" effect is probably the dominant one. In fact, this smallest GNB also features a significant truncation percentage of 38% (see Table S1), defined as $(1 - h/h')$ in %, and thus possesses a blunt tip, this minimizing the tip effect independently on the tips shape (spherical or flat). Overall, this explains why, in this case, the BCS's absorption is blue-shifts with respect to the BCF one. Conversely, as the $AR$ and the base size are slightly increased (i.e., to $AR$ = 2.40 and $w$ = 15 nm) and the truncation percentage reduced to 27%, the BCS absorption is red-shifted with respect to the BCF model. This result indicates that, in this case, the volume modification has a smaller contribution than in the previous case (reducing the LSPR blue-shift) and the presence of a sharpened tip determines a sizable tip effect (and thus a red-shift) that is clearly larger for the sharp spherical tip than for the planar flat tip. This trend is essentially maintained if the $AR$ and the base size are further increased and the truncation percentage further reduced, but the BCS vs BCF absorption difference tends to shrink as a consequence of the reducing size of the tip with the elongation of the GNB particle, which will make fading the importance of the choice of the tip shape as the $AR$ increases, see Figure 2a.

The BCS model thus provides the best agreement with experimental evidence and hereafter will be used as the reference shape for all remaining simulations. In Figure 2b, the experimental (normalized) absorption spectra of the GNBs at different $AR$s are compared with BEM simulations using the BCS reference shape. The experimental plasmon resonances are, as expected, strongly prominent over the interband transitions of bulk gold lying at shorter wavelengths (i.e., in the UV spectral window not shown in Figure 2b) while weaker absorptions are found at wavelengths in the 500−600 nm range. Our results suggest that in this spectral region several contributions give rise to such absorptions: (i) the presence of spherical spheres (resulting as byproduct of the GNBs synthesis); (ii) the transverse plasmon resonances due to the interacting light polarized along the short axis of GNBs; and (iii) multipolar resonance absorption peaks. As shown in Figure S4 in the SI, we checked the effects of the polarization of the exciting electromagnetic field over the plasmonic resonances, and we found that, for all GNBs here considered, the transverse resonances are located at 506 nm. Moreover, for the largest GNB ($AR$ = 3.1) an additional absorption peak at around 600 nm due to multipolar resonance can be detected in the simulated spectrum (which disregard contributions from spheres and transverse resonances), suggesting that part of the absorption detected experimentally for this large, elongated GNB, feature such of a contribution. Notably, experimental spectra of even more elongated GNBs than that with $AR$ = 3.1 reported in ref 21 feature clearer signatures of this multipolar contribution, at correspondingly more red-shifted wavelenghts. Finally, as shown in Figure S3, the intensities of the plasmonic bands depend on the volume of the GNBs, i.e., the greater the volume of the particle the more intense is its absorption band, in line with what is known for simpler shapes.

**Base Size and Truncation Effects.** Given the good agreement between experimental and theoretical LSPR spectra computed using the BEM method for the BSC model, with better accuracy than previously reported DDA computations, we performed simulations on a larger set of synthetic GNBs featuring a wider range of base sizes (up to ca. 37 nm), $AR$ and $AR_{Id}$ (low to 1.84 and 3.16, respectively) and tip truncation percentages (up to 48%), as reported in Figure 3 and Table S2

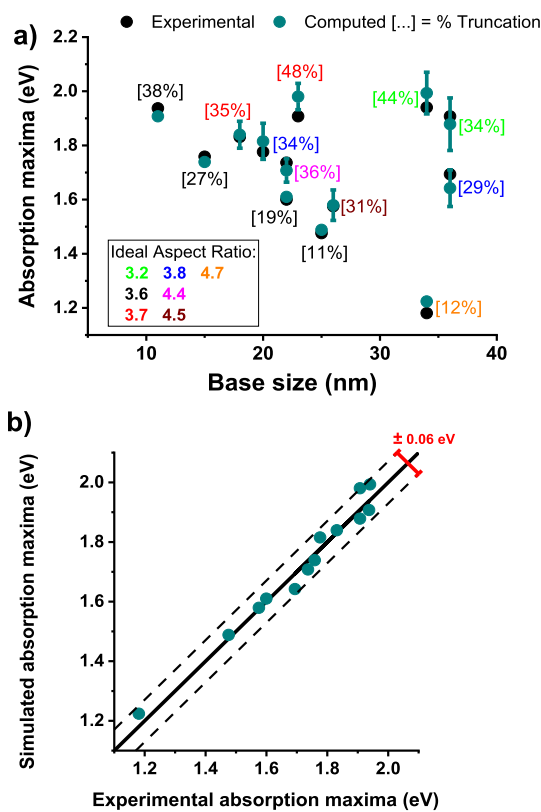

Figure 3. (a) Comparison of experimental (black solid circles) LSPR absorption maxima and theoretical values (green solid circles), computed with average single-particle BCS models, for all the GNBs considered. These GNBs feature a large geometrical variability, including different base sizes and various $AR_{Id}$s (in colored text) and truncation percentages (in square brackets). For the GNBs TEM images analyzed in this work, the largest difference ($\xi$) in LSPR energies, with respect to the average single-particle computations, are reported as ($\pm\xi$) error bars of theoretical values. (b) Correlation plot of experimental vs simulated absorption maxima.

in the SI. As reported in Table 1, the LSPR absorption maxima computed, within the full-EM framework, with the BEM approach and BCS geometrical model feature overall a good agreement with all experimental data, with energy difference discrepancies always below 0.06 eV (see Figure 3b). The simulated LSPR absorptions are computed using an average GNB particle, i.e. averaging the geometrical parameters, i.e. height $\langle h \rangle$, aspect ratios $\langle AR \rangle$ and $\langle AR_{Id} \rangle$, that can be extracted

Table 1. Experimental (in nm) and Computed (in nm and eV) LSPR Absorption Maxima for All GNBs Samples[a]

| exp. LSPR (nm) | theory LSPR (nm) | theory LSPR (eV) | $\Delta E$ (eV) (Exp−Comp) | geom. var. $\xi$ (eV) | $W$ (nm) |
|---|---|---|---|---|---|
| 639[b] | 622 | 1.993 | −0.05 | 0.08 | 34.3 |
| 650[b] | 660 | 1.878 | 0.03 | 0.1 | 36.7 |
| 650[b] | 626 | 1.981 | −0.07 | 0.05 | 23.0 |
| 677[c] | 674 | 1.839 | −0.008 | 0.05 | 18.1 |
| 698[c] | 683 | 1.815 | −0.04 | 0.07 | 20.1 |
| 714[b] | 726 | 1.708 | 0.03 | 0.04 | 21.9 |
| 732[c] | 755 | 1.642 | 0.05 | 0.07 | 36.3 |
| 787[b] | 785 | 1.579 | −0.004 | 0.06 | 26.3 |

[a]Theoretical values are computed with BEM method and BSC morphology, using a single GNB particle model based on the average of geometrical parameters (reported in Table S2 in the SI) extracted from experimental TEM images of each GNB sample. Differences between experimental and computed values are reported as $\Delta E$, in eV. Geometrically-induced maximum variations on the computed LSPR energies ($\xi$, in eV) observed for the ensembles of GNBs in each sample are also reported. [b]Data from ref 21. [c]Data from ref 20.

from the experimental TEM images, as described in the Computational Details section.

In order to assess the uncertainty of the computed LSPR associated with the geometrical variance of the synthetic GNBs, we analyzed TEM images from the samples reported in refs 20 and 21. We used the standard deviations ($\pm 1\sigma$) of the three geometrical parameters $\langle w \rangle$, $\langle AR \rangle$, and $\langle AR_{Id} \rangle$ (see the Computational Details section) to build eight ($2^3$) GNBs' "extreme cases" models for each sample and we computed their individual LSPR absorption maxima to have a rough estimate of their variation as a function of these parameters. In Figure 3a we report the largest difference ($\xi$) in LSPR energy with respect to the average GNBs as ($\pm \xi$) error bars, showing that the geometrical uncertainty derived from the extraction of the GNBs' parameters from TEM images is significantly large at already $\pm 1\sigma$ from the average and generally even larger than the difference between the simulated and experimental LSPR absorption maxima. This fact suggests that the sensitivity and accuracy of the BEM approach is sufficiently high that the results of the simulations are mostly affected by the choice of the NP extracted from the TEM image and from the corresponding geometrical parameters. This outcome highlights the importance of averaging all of the geometrical parameters extracted from TEM images rather than selecting one random single NP from the sample. Finally, we tested how much the result of such protocol differs from the explicit computations of all the LSPR absorption maxima of any NP present in the sample. Since these explicit ensemble computations have a computational cost significantly larger than a (averaged) single particle computation, we performed the simulations for only one representative sample. In particular, we selected the GNB sample featuring both a large geometrically induced variation on the LSPR energy (i.e., a large $\xi$) and one of the largest discrepancy between the experimental LSPR maximum and the single-particle theoretical value (i.e., a "large" $\Delta E$(Exp−Comp)). We have thus computed the LSPR spectrum of the full ensemble of the GNB featuring an experimental LSPR maximum at 639 nm (i.e., 1.940 eV), featuring $\xi = 0.08$ eV and $\Delta E$(Exp−Comp) = −0.05 eV, and compared it with the spectrum of the (averaged) single-particle, showing a deviation of LSPR maximum of only 0.01 eV and an almost identical LSPR spectra (with obviously a broader line shape in the full ensemble spectrum), as reported in Figures S1 and S2 in the SI.

**Base Size Limits for QSA Computations.** After validating BEM approach to EM scattering for GNBs particles with broad geometrical variety in terms of base size, $AR_{Id}$, and tip truncation (i.e., $AR$), we investigated how well QSA results could compare against full-EM computations. The assessment of QSA computations, in fact, is crucial to allow theoretical estimation of interactions between GNBs and molecular transition dipole moments in nanohybrids materials. Figure 4 reports the comparison of the experimental and full-EM simulations data described in section Assessment of GNBs' model shapes with the QSA estimates of the LSPR absorption maxima using the BSC model. The outcome shows good agreement between the QSA results and the full-EM (and the experimental) data. As expected from the intrinsic limitation of the QSA approach (see the Introduction), the accordance worsens while increasing the GNBs' particle size, which (in the reported cases) is associated with an increase of the $AR$.

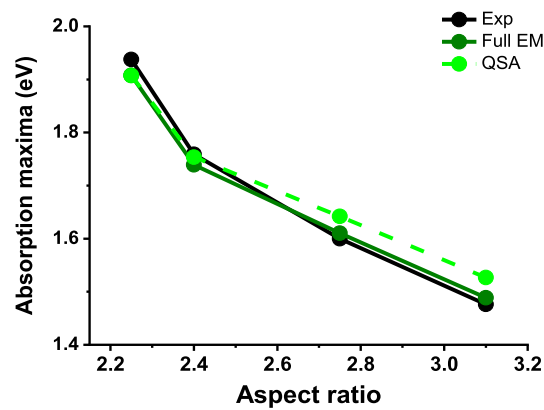

**Figure 4.** Comparison of experimental (black solid circles) LSPR absorption maxima and theoretical values for the BCS models computed with full-EM (dark green solid circles) and QSA (light green solid circles) methods.

In order to better evaluate the error of the QSA calculations as a function of the GNBs' size, we built a series of GNB models with increasing base size $w$, keeping the $AR$ and the $AR_{Id}$ fixed and then we compared the LSPR absorption maxima computed with the QSA and the full-EM approaches. The outcome of these computations is reported in Figure 5a and it shows how the particle size-independent of QSA estimates of the LSPR absorption maxima converge to the full-EM results for decreasing GNBs' base sizes.

As previously mentioned, the full-EM computations show how the increase of the particle base size (that, for a given $AR$ and $AR_{Id}$, corresponds to an increase of particle volume) is associated with a red-shift of the LSPR absorption, limiting the accuracy of the QSA computations to a finite range of base sizes, which depends on the chosen $AR$ and $AR_{Id}$. Thus, in order to evaluate the extension of this range we computed the error (in percentage) in QSA absorption energies with respect to full-EM results as a function of the base size, for each $AR$ and for a fixed $AR_{Id}$ (equal to 3.6, see Table S1). As shown in Figure 5b, to stay within an error of 5% in LSPR absorption energy, the maximum base size should be lower than 36 nm for this class of GNBs. Considering that the typical base sizes of GNB objects are below 40 nm (as all GNBs investigated here, see Tables S1 and S2 in the SI) and that $AR_{Id}$s smaller than the one selected (that eventually would slightly reduce the maximum base size value) are not quite common, our result definitively represents a good estimation of the typical base size limit that one can reach one modeling GNBs LSPR absorption spectra within the QSA methodology.

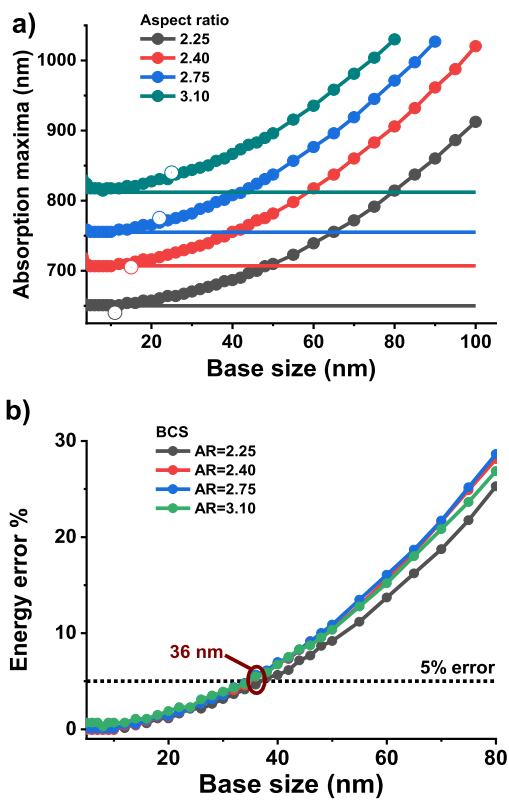

**Figure 5.** (a) Comparison between full-EM (colored solid circles) and QSA calculations (colored flat lines) of LSPR absorption maxima, as a function of the GNBs base sizes and for different *ARs* (color coded from 2.25 to 3.10) and fixed $AR_{Id}$ (equal to 3.6), with empty colored circles reporting the experimental data (at specific base sizes). (b) Energy errors (in percentage) on the LSPR absorptions computed with QSA method with respect to the full-EM reference values, as a function of the GNBs' base size, for different *ARs* (color coded from 2.25 to 3.10) and fixed $AR_{Id}$ (equal to 3.6).

## ■ CONCLUSIONS

In this work we proved that classical electrodynamics full-EM scattering calculations, solved numerically employing BEM approaches, can accurately describe the LSPR absorption properties of synthetic GNBs' samples, if biconical models with spherical tips are employed. BEM simulations are sensitive enough to the GNBs' geometries that can provide results fully consistent with experimental evidence and with higher accuracy than DDA computations. The GNBs' models can be constructed by extracting geometrical parameters from TEM images of synthetic samples, using a single (random) representative particle, the full set of particles, or an average single-particle obtained by averaging the geometrical parameters over the ensemble of particles. Our results demonstrate that the latter option provides results very similar to the full-ensemble computations, with a gain in computational cost, and suggest that random selection of a single nanoparticle from TEM images to model a GNB could entail a large error on the computed LSPR absorption. Finally, the performances of QSA simulations of GNBs' LSPR absorption spectra have been compared to the accurate full-EM approach, since reliable QSA computations would allow appropriate characterization of the interactions between molecular electron densities and gold nanoparticles that occur in GNB-based organic nanohybrids. Despite QSA computations of LSPR absorption energies are not sensitive to the particle sizes, they proved to be in good agreement with full-EM calculations for an extended range of GNBs' base sizes, proving the applicability of this approach to synthetic GNBs and paving the way to the study of the optical properties of GNB-based nanohybrid materials.

## ■ ASSOCIATED CONTENT

### ⓢ Supporting Information

The Supporting Information is available free of charge at https://pubs.acs.org/doi/10.1021/acs.jctc.0c00269.

Additional results as shown in Figures S1−S4, Tables S1 and S2, and refs 1 and 2 (PDF)

## ■ AUTHOR INFORMATION

### Corresponding Authors

**Stefano Baroni** − *SISSA—Scuola Internazionale Superiore di Studi Avanzati, 34136 Trieste, Italy*; Email: baroni@sissa.it

**Stefano Corni** − *Dipartimento di Scienze Chimiche, Università di Padova, 35131 Padova, Italy; Istituto di Nanoscienze, Consiglio Nazionale delle Ricerche CNR-NANO, 41125 Modena, Italy;* ⓞ orcid.org/0000-0001-6707-108X; Email: stefano.corni@unipd.it

**Marco Garavelli** − *Dipartimento di Chimica Industriale "Toso Montanari", Università degli Studi di Bologna, I-40136 Bologna, Italy;* ⓞ orcid.org/0000-0002-0796-289X; Email: marco.garavelli@unibo.it

**Ivan Rivalta** − *Laboratoire de Chimie UMR 5182, CNRS, Université Lyon 1, Univ Lyon, Ens de Lyon, F-69342 Lyon, France; Dipartimento di Chimica Industriale "Toso Montanari", Università degli Studi di Bologna, I-40136 Bologna, Italy;* ⓞ orcid.org/0000-0002-1208-602X; Email: i.rivalta@unibo.it

### Authors

**Jacopo Marcheselli** − *SISSA—Scuola Internazionale Superiore di Studi Avanzati, 34136 Trieste, Italy*

**Denis Chateau** − *Laboratoire de Chimie UMR 5182, CNRS, Université Lyon 1, Univ Lyon, Ens de Lyon, F-69342 Lyon, France*

**Frederic Lerouge** − *Laboratoire de Chimie UMR 5182, CNRS, Université Lyon 1, Univ Lyon, Ens de Lyon, F-69342 Lyon, France;* ⓞ orcid.org/0000-0003-2909-527X

**Patrice Baldeck** − *Laboratoire de Chimie UMR 5182, CNRS, Université Lyon 1, Univ Lyon, Ens de Lyon, F-69342 Lyon, France*

**Chantal Andraud** − *Laboratoire de Chimie UMR 5182, CNRS, Université Lyon 1, Univ Lyon, Ens de Lyon, F-69342 Lyon, France*

**Stephane Parola** − *Laboratoire de Chimie UMR 5182, CNRS, Université Lyon 1, Univ Lyon, Ens de Lyon, F-69342 Lyon, France*

Complete contact information is available at:
https://pubs.acs.org/10.1021/acs.jctc.0c00269

**Notes**
The authors declare no competing financial interest.

# ACKNOWLEDGMENTS

I.R., M.G., F.L., P.B., C.A., and S.P. thank ENS Lyon (France) and the financial support from the LABEX iMUST (ANR-10-LABX-0064) of Université de Lyon, within the program "Investissements d'Avenir" (ANR-11-IDEX-0007) operated by the French National Research Agency (ANR). S.C. acknowledge funding from the ERC under the Grant ERC-CoG-681285 TAME-Plasmons. I.R. acknowledges the use of HPC resources of the "Pôle Scientifique de Modélisation Numérique" at the ENS Lyon, France.